# STM-UNet: An Efficient U-shaped Architecture Based on Swin Transformer and Multi-scale MLP for Medical Image Segmentation

Lei Shi[1,2], Tianyu Gao[1], Zheng Zhang[1] and Junxing Zhang[1, *], *IEEE Member*
[1] College of Computer Science, Inner Mongolia University, Hohhot, China.
[2] Baotou Medical College, Baotou, China.
Email: shilei@mail.imu.edu.cn; junxing@imu.edu.cn

*Abstract*—Automated medical image segmentation can assist doctors to diagnose faster and more accurate. Deep learning based models for medical image segmentation have made great progress in recent years. However, the existing models fail to effectively leverage Transformer and MLP for improving U-shaped architecture efficiently. In addition, the multi-scale features of the MLP have not been fully extracted in the bottleneck of U-shaped architecture. In this paper, we propose an efficient U-shaped architecture based on Swin Transformer and multi-scale MLP, namely STM-UNet. Specifically, the Swin Transformer block is added to skip connection of STM-UNet in form of residual connection, which can enhance the modeling ability of global features and long-range dependency. Meanwhile, a novel PCAS-MLP with parallel convolution module is designed and placed into the bottleneck of our architecture to contribute to the improvement of segmentation performance. The experimental results on ISIC 2016 and ISIC 2018 demonstrate the effectiveness of our proposed method. Our method also outperforms several state-of-the-art methods in terms of IoU and Dice. Our method has achieved a better trade-off between high segmentation accuracy and low model complexity.

## I. INTRODUCTION

Medical image segmentation is an important task in medical image analysis. Accurate segmentation of lesion size and morphology is useful for determining the grade of the disease, as well as guiding the pre-surgical analysis and the next treatment plan. Automated segmentation of medical images based on deep learning can assist doctors to make more accurate diagnoses and speed up the first consultation. Given the characteristics of medical images, the U-Net [1] based convolutional neural network model has long ruled the medical image segmentation field, becoming the de facto standard since the year of 2015. Since the vision Transformers [2, 3] were proposed, researchers have attempted to combine the vision Transformer with the U-shaped structure. Medical image segmentation architectures based on CNN with Transformer as well as pure Transformer were proposed to take advantage of the Transformer's priority in modeling global features and long-range dependency and further improve segmentation performance. In addition, a few works [4] have also attempted to introduce MLP into U-shaped architecture to improve the performance of medical image segmentation or realize a lightweight model.

However, limitations and challenges remain. Firstly, the existing models fail to effectively combine the respective advantages of CNN, Transformer, and MLP, and do not place the above modules in the most suitable position in the U-shaped architecture to organically improve the segmentation performance; Secondly, although the existing work tries to use MLP at the bottom of the U-shaped architecture, it fails to effectively extract the multi-scale features of the MLP, thus the classification ability of MLP is not fully exploited; Thirdly, the model complexity of most existing architecture based on CNN and Transformer is generally high, which is not conducive to be deployed on mobile devices for training or inference. In many specific tasks (e.g., skin lesion segmentation), increasing the model complexity does not lead to further improvement in segmentation accuracy.

To address the above problems and challenges, in this paper, we propose a segmentation model called "STM-UNet", which leverages conventional CNN for encoder and decoder to extract and recover image features. Meanwhile, in the skip connection of U-shaped architecture, we add the Swin Transformer block [3] in form of residual connection to fuse the local features and global features of each layer, so that the overall features of the lesion region can be recovered effectively in the decoder side; we have proposed a model called PCAS-MLP and incorporate it into the bottleneck of U-shaped structure with richest classification information. The novel module adds a parallel convolution module into AS-MLP [5], which effectively improves the classification ability of pixels and makes a contribution to the improvement of overall segmentation accuracy. Our proposed method not only has high segmentation accuracy but also remains lightweight compared to most of state-of-the-art methods. STM-UNet is still suitable for deployment in mobile devices and scenarios. Experimental results on two public datasets show that our method outperforms state-of-the-art methods in terms of IoU and Dice, confirming the effectiveness and advancement of the proposed method.

The main contributions of this paper are as follows:

- The Swin Transformer block is added to skip connection of the U-shaped architecture in form of residual connection, which can effectively fuse the global features and local features of each layer and enhance the modeling ability of global features and long-range dependency.

- A parallel convolution module is added to AS-MLP in the proposed architecture, which can effectively extract the

This work was partially supported by the National Natural Science Foundation of China (Grant No. 61261019), the Inner Mongolia Science & Technology Plan (Grant No. 201802027), and the Inner Mongolia Autonomous Region Natural Science Foundation (Grant No. 2018MS06023).
* Corresponding author: Junxing Zhang (junxing@imu.edu.cn).

multi-scale features of MLP, and then improve the segmentation performance;

• The segmentation performance of the proposed architecture outperforms several CNN, Transformer, and MLP based baselines on two publicly available datasets, demonstrating the effectiveness of our proposed method.

## II. RELATED WORK

Before the advent of vision Transformers, U-shaped architectures based on convolutional neural networks dominated the deep learning-based medical image segmentation methods. For example, Unet++[6] leverages a nested and dense design for the skip connection based on U-Net to better capture features at each level and fuse them efficiently. Unet3+[7] further improves Unet++ with a full-scale skip connection, which includes not only encoder-decoder inter-connections but also decoder-decoder inner connections. However, due to the intrinsic inductive bias, CNN can only focus on local features of images and it cannot model long-range dependency and global features, which restricts the further improvement of medical image segmentation accuracy.

Recently, inspired by the success of vision Transformer in natural images [2] [3], some work attempted to apply vision Transformer to medical image segmentation tasks and incorporate it into U-shaped architecture. TransUNet [8] is the first public work combining Transformer with medical image segmentation. It designs a hybrid structure of CNN and Transformer as an encoder: CNN is used for feature extraction and the extracted feature map is fed into a standard Transformer. Hu Cao et al. [9] proposed a pure Transformer architecture called Swin-Unet, which mainly consists of an encoder, bottleneck, and decoder. The Swin Transformer block is the main component in all above three parts. The input image is sliced into patches and linear embedded, and then fed into the encoder of full Transformer. The Swin Transformer block in the decoder is symmetrical to the counterpart of encoder and is skip-connected in the horizontal direction, thus enabling features from different levels to incorporate in final prediction map. UCTransNet [10] mainly pays attention to improving the skip connection of U-Net by embedding a channel Transformer module. The channel Transformer module consists of two sub-modules: the multi-scale Channel Cross fusion with Transformer (CCT) and the Channelwise Cross-Attention (CCA).

In recent years, the improved MLPs [5] [11] [12] have been back to computer vision classification tasks and achieved comparable performance to CNN and Transformer. Inspired by this, UNeXt [4] placed the MLP in the bottleneck of U-shaped architecture for the first time to implement a lightweight segmentation model.

However, the existing architectures fail to combine the respective advantages of CNN, Transformer, and MLP. They have not fully exploited the potential of using MLP in U-Net as well. The model proposed in this paper effectively incorporates CNN, Swin Transformer, and MLP in one architecture, while proposing a novel PCAS-MLP module. This design enables the model to focus on both local and global features of lesions, while enhancing the classification ability of MLP, thus effectively improving medical image segmentation performance in general.

## III. PROPOSED METHOD

### A. The architecture design

The proposed architecture of this paper is shown in Fig.1. Similar to the vanilla U-Net, the architecture consists of encoder, decoder, bottleneck, and skip connection. The encoder part adopts the traditional convolutional module, which consists of 3*3 convolution, maxpooling, and activation function. The role of the encoder is to implement downsampling in a convolutional manner and focus on extracting local features of the original image layer by layer. The decoder part leverages convolution and bilinear interpolation to enlarge both width and height of the input feature. The role of decoder is to restore the feature map of the bottleneck to the same size as the input image layer by layer. In the bottleneck, we add the newly designed PCAS-MLP module. It is an improvement of AS-MLP by adding the parallel convolution module to extract the multi-scale features of the MLP. In the skip connection, we add Swin Transformer blocks to enhance the modeling ability of global features and long-range dependency for the proposed architecture.

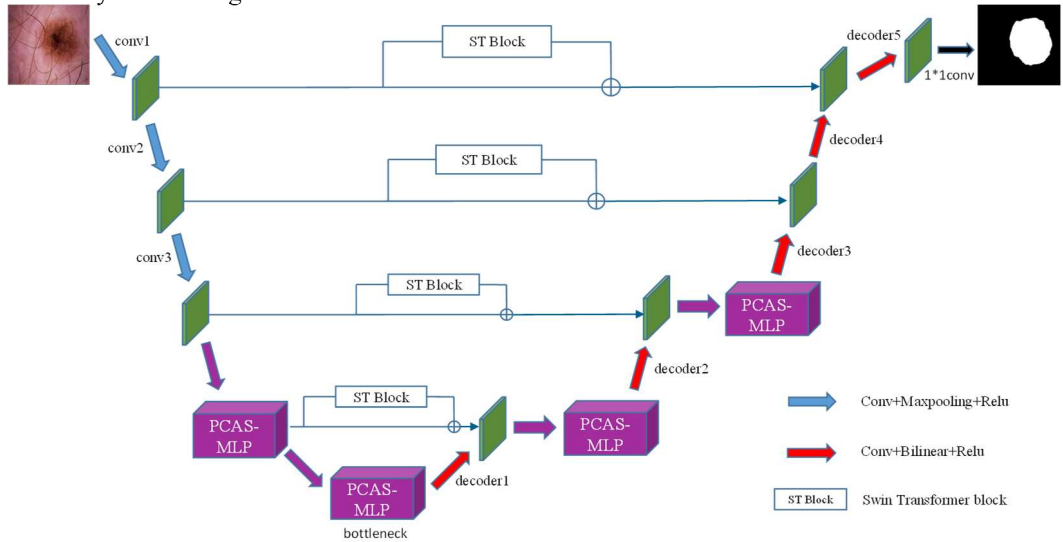

Fig. 1. The overall architecture of our proposed method. The network consists of a mixture of convolution, Transformer, and MLP. The convolution modules are mainly present in the encoder and decoder, and the PCAS-MLP is placed at the bottom of the U-shaped architecture for better extraction of the classification information of pixels. The Swin Transformer block is added to skip connection in form of residual connection to better extract the global feature of the feature map in each layer. The convolution, Transformer, and MLP are organically combined to enable the network to fully extract various beneficial features in the lesion region of medical images, thus effectively improving segmentation performance.

## B. Residual Swin Transformer block

Swin Transformer [3] has been extremely successful in natural image processing and is one of the most powerful vision transformer backbones in performance. The skip connection in this paper incorporates the Swin Transformer block and is deployed in a residual way. The original skip connection of U-Net only fuses the encoder with decoder at the same layer by adding or concatenation. However, the inductive bias property of convolution prevents the original skip connection from focusing on the global features of each layer. To this end, we add Swin Transformer block to skip connection to extract global features of each layer and fuse them with local features extracted by CNN in an additive manner. Thus, the encoder information recovered from each layer would contain both local features and global features to further improve the segmentation performance. It is worth noting that the Swin Transformer block used in this paper does not require pre-training. The basic structure of two consecutive Swin Transformer blocks is shown in Fig.2. The main difference between Swin Transformer and standard Transformer is the replacement of multi-head self-attention (MSA) with regular windowing MSA (W-MSA) and shifted windowing MSA (SW-MSA). The design of the shifted window enables the interaction across windows, and thus better extracts global features of the image. In Fig.2, $\hat{z}^l$ and $\hat{z}^{l+1}$ are the output features of W-MSA and SW-MSA, respectively. $z^l$ and $z^{l+1}$ are the output features of MLP in each block. The successive Swin Transformer blocks can be computed as follows:

$$\hat{z}^l = W-MSA(LN(z^{l-1})) + z^{l-1},$$
$$z^l = MLP(LN(\hat{z}^l)) + \hat{z}^l,$$
$$\hat{z}^{l+1} = SW-MSA(LN(z^l)) + z^l,$$
$$z^{l+1} = MLP(LN(\hat{z}^{l+1})) + \hat{z}^{l+1} \quad (1)$$

In this paper, the parameters of Swin Transformer blocks are adjusted to be suitable for the characteristics of the used datasets.

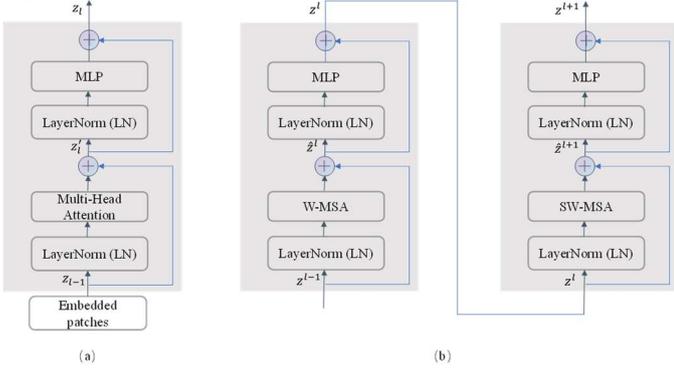

Fig. 2. The basic structure of consecutive Swin Transformer blocks: (a) a standard Transformer block; (b) two successive Swin Transformer blocks.

## C. PCAS-MLP

MLP is naturally suitable for image classification, especially since recent works [11-12] have achieved classification performance comparable to CNN and Transformer on ImageNet and other datasets. Meanwhile, the bottleneck of U-shaped architecture is usually rich in high-level information related to the class of individual pixels. Therefore, it is most appropriate to incorporate MLP into the bottleneck of U-shaped architecture. Here we adopt the core idea of AS-MLP (Axial Shifted MLP) [5] to extract the information from feature maps in different axis. axial shift mainly contains horizontal shift and vertical shift.

Fig.3(a) shows the input feature map with the shape of B*C*H*W. To facilitate the analysis of horizontal shift, we ignore batch size and height, and assume C=5 and W=5. Supposing shift size=5, the input feature map is first cut into 5 blocks along the channel direction. And then, each block is shifted along the horizontal direction respectively by a stride of [-2, -1, 0, +1, +2], in which "-" denotes shifting to the left, "+" denotes shifting to the right, and the empty positions will be padded zero. Finally, the feature map in the red dashed box in Fig.3(b) will be taken and passed to the next fully connected layer. The principle of vertical shift is similar, except that the cut blocks are shifted along the vertical direction.

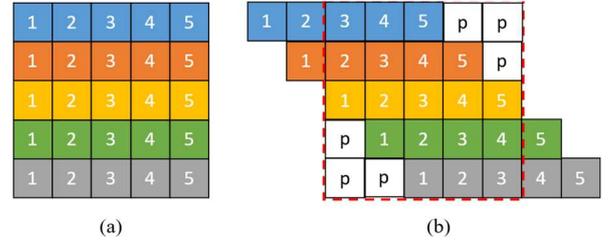

Fig. 3. Horizontal shift (shift size=5).

The experimental results show that the AS-MLP with a serial type performs a better segmentation performance for the U-shaped architecture. As a result, we use a serial type of AS-MLP instead of a parallel one, namely, we first shift the feature map in the height direction and then in the width direction.

Adding convolution into MLP can encode position information effectively and improve the classification performance of MLP [4][13]. UNext also uses AS-MLP as the bottleneck of the U-shaped architecture, but merely adding 3*3 convolution to AS-MLP is not enough to extract the multi-scale features of MLP. For this reason, inspired by Inception serial modules [14,15], we add a parallel convolution module to the AS-MLP to extract the multi-scale features of the module after the first fully connected layer. Fig.4 shows the basic structure of PCAS-MLP (Parallel Convolution AS-MLP). A parallel convolution module is added between fc1 and horizontal shift. This module consists of three convolutions with kernel sizes of 1*1, 3*3, and 5*5 in a parallel manner. The combination of different-sized convolutions can fully extract the multi-scale information of the input feature map, thus improving the classification performance of MLP and eventually segmentation performance of the whole network. The parallel convolution module differs from Inception-v1 in two aspects: firstly, there is no 3*3 max-pooling in our module; secondly, the way of fusing different convolutions takes adding rather than concatenation. The parallel convolution module designed is a novel and useful module that can be organically integrated with MLP to improve segmentation performance.

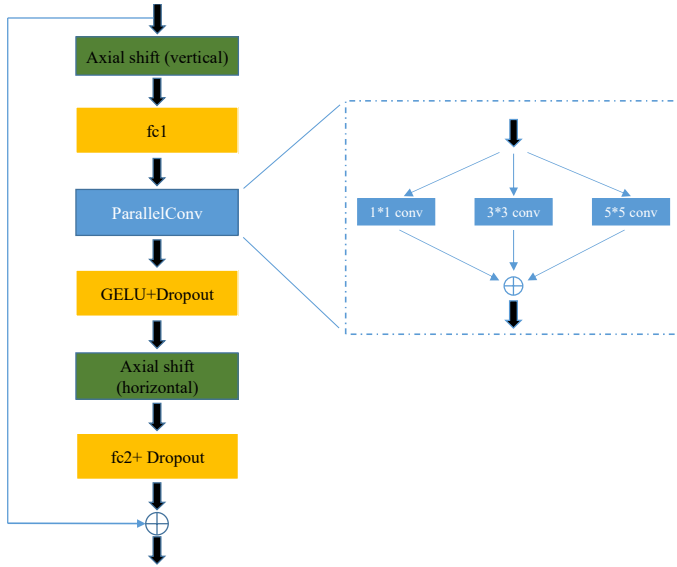

Fig. 4. The basic structure of PCAS-MLP, where the parallel convolution module is added between fc1 and horizontal shift.

IV. EXPERIMENTS AND RESULTS

*A. Datasets*

Two datasets are used in this paper to evaluate the performance of the proposed method and compare it with other SOTA methods. The ISIC 2016 dataset [16] and ISIC 2018 dataset [17,18] were all released by ISIC Challenge, for evaluating different automated segmentation methods on skin lesion. The ISIC 2016 dataset contains 900 training images and 379 test images. The ISIC 2018 dataset contains 2594 training images and 1000 test images. For ISIC 2018 dataset, previous works always performed 5-fold cross-validation on the training set or directly repartitioned the training set. Recently, The ISIC Challenge released the ground truths corresponding to the 1,000 test images for ISIC 2018 Task 1. As a result, for both datasets, we split the train set with an 8:2 proportion to train and valid, and then evaluate the segmentation performance on the official test set. All the images and masks are resized to the resolution of 512*512.

*B. Implementation Details*

In our experiments, we trained our model and other SOTA models for 300 epochs with an initial learning rate of 0.0001. Adam is adopted as the optimizer and we set the training batch size to 8. For testing, the batch size is set to 1 to ensure the uniqueness of test results. An NVIDIA A100 is used for all our experiments and Pytorch is used as the deep learning framework. For fair comparison, we basically followed UNext [4] for other settings that are not mentioned here.

*C. Evaluation Metrics*

To evaluate the segmentation performance of different methods, we adopt the dice coefficient (Dice) and intersection over union (IoU) to measure the difference between the predicted maps and the ground truths. We calculate the IoU and Dice of each image, and average the whole testset to get the value of mIoU and mDice.

*D. Comparison with state-of-the-art methods*

To verify the effectiveness of the proposed method in this paper, we compare the segmentation performance of STM-UNet with several state-of-the-art methods. Table 1 shows the quantitative results evaluated on the test set of different datasets. We choose UNet [1], UNet++ [6], UNet3+ [7], UCTransNet [10] and UNeXt-L [4] as the baselines for comparison. UNet, UNet++, and UNet3+ are pure CNN architectures, [8] and [10] are hybrid architectures based on Transformer and CNN, and UNeXt-L is a hybrid architecture based on MLP and CNN. For fair comparison, we set the input channel of each layer for all U-shaped architectures to [32, 64, 128, 256, 512]. Note that we do not use any pretraining in all our experiments.

TABLE I. QUANTITATIVE RESULTS EVALUATED ON TEST SET OF DIFFERENT DATASETS

| Methods | ISIC2018 | | ISIC2016 | | Params(M) |
|---|---|---|---|---|---|
| | mIoU | mDice | mIoU | mDice | |
| UNet [1] | 0.7507 | 0.8386 | 0.8315 | 0.8984 | 9.85 |
| UNet++ [6] | 0.7526 | 0.8409 | 0.8212 | 0.8889 | 11.8 |
| UNet3+ [7] | 0.7400 | 0.8305 | 0.8243 | 0.8909 | 6.75 |
| UCTransNet [10] | 0.7891 | 0.8672 | 0.8436 | 0.9050 | 17.07 |
| UNeXt-L [4] | 0.7843 | 0.8671 | 0.8414 | 0.9058 | 3.99 |
| STM-Unet (Ours) | **0.7984** | **0.8751** | **0.8463** | **0.9094** | 6.12 |

From Table 1, it can be seen that the segmentation performance of our method is the highest for both ISIC2016 and ISIC2018. For example, our method outperforms the second place in mIoU and mDice by 0.93% and 0.79%, respectively for ISIC2018. It is worth noting that the Params (M) and FLOPs (G) of Our STM-UNet are only slightly higher than UNeXt-L, but lower than other compared methods. From Table 1, we also find that the complex design of architectures does not lead to improvement of segmentation performance for certain tasks. The reason may lie in the region of skin lesion always varies in the whole image, and it is difficult to extract all the features of the lesion by simply applying a more complex, especially purely CNN-based architecture. Therefore, the architecture proposed in this paper can guarantee high segmentation performance while maintaining relatively low complexity, and can still be applied to mobile devices as a lightweight model.

Fig.5 shows the qualitative segmentation results of several dermatological images using different methods. As seen in Fig.5, the lesion area predicted using our method is the closest to the ground truth. In addition, our method has the smallest predicted area of FP (false positive) among all methods, avoiding unnecessary examination treatment and psychological stress to the patients. The analysis of the qualitative results shows that our method can produce more accurate automatic segmentation results due to the focus on modeling long-range dependency and the enhanced classification capability of pixels.

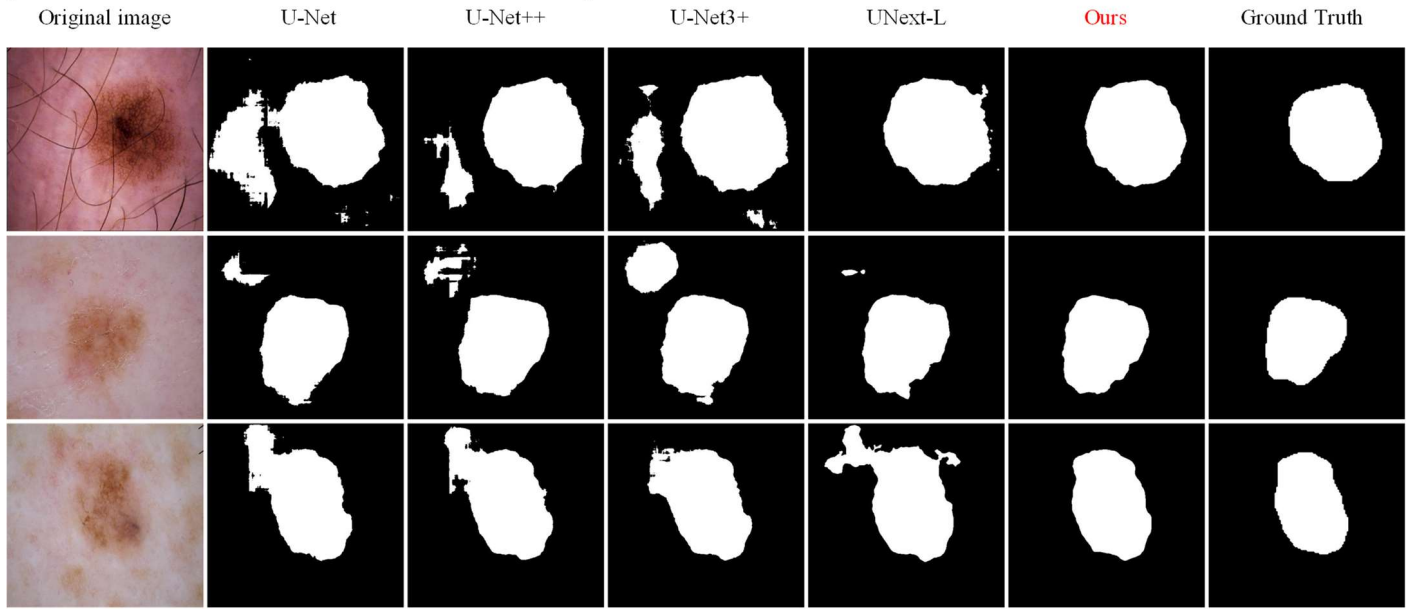

Fig. 5. Visualization of segmentation results with different methods

*E. Ablation study*

We perform the ablation study on each added part in the proposed architecture to evaluate the impact of each part on the segmentation performance. Table 2 shows the results of ablation experiments on the test set of ISIC2018. We use vanilla UNet with AS-MLP in bottleneck as the baseline. We can see from Table 2 that the addition of both Swin Transformer block and ParallelConv has a positive effect on improving the segmentation performance of the baseline, demonstrating that each added part contributes independently to the final segmentation results. This is because the Swin Transformer block enhances the extraction ability of global features of the lesion, while ParallelConv extracts the multi-scale features of the MLP, thus improving the classification accuracy of pixels in the prediction map. Ultimately, the STM-UNet with the addition of both Swin Transformer block and ParallelConv has the best segmentation results, indicating that aggregating the advantages of each module can jointly improve the segmentation performance.

TABLE II. ABLATION STUDY ON ISIC2018 TEST SET

| ST block | ParallelConv | ISIC2018 | |
|---|---|---|---|
| | | mIoU | mDice |
| × | × | 0.7720 | 0.8586 |
| √ | × | 0.7903 | 0.8667 |
| √ | √ | 0.7984 | 0.8751 |

## V. CONCLUSION

In this paper, we have proposed a novel and efficient architecture, namely STM-UNet, for medical image segmentation. We add Swin Transformer blocks into the skip connection of U-shaped architecture to enhance the modeling ability of global features and long-range dependency. In addition, we propose a novel module called" PCAS-MLP", which adds a parallel convolution module in axial-shifted MLP to extract muti-scale features and enhance the ability to classify pixels for MLP. We conduct extensive experiments on two public datasets to evaluate our proposed method. Experimental results show that our method outperforms several state-of-the-art methods in terms of IoU and Dice. Our method aims to achieve a better trade-off between segmentation performance and model complexity. Our code will be available after the formal publication of this paper.